\documentstyle[12pt,epsf]{article}
\newcommand{\beq}{\begin{equation}}
\newcommand{\eeq}{\end{equation}}
\newcommand{\bdis}{\begin{displaymath}}
\newcommand{\edis}{\end{displaymath}}
\newcommand{\bea}{\begin{eqnarray}}
\newcommand{\eea}{\end{eqnarray}}
\newcommand{\barr}{\begin{array}}
\newcommand{\earr}{\end{array}}
\title{Dynamically Driven Renormalization Group}
\author{Alessandro Vespignani$^{(1)}$, Stefano Zapperi$^{(2)}$ 
and Vittorio Loreto$^{(3)}$}
\begin{document}
\baselineskip 18pt
\pagestyle{myheadings}
\markboth{Vespignani, Zapperi and Loreto}{Vespignani, Zapperi and Loreto}
\maketitle
\centerline{1) {\em Instituut-Lorentz, University of Leiden, P.O. Box 9506}}
\centerline{{\em 2300 RA, Leiden, The Netherlands}}
\centerline{2) {\em Center for Polymer Studies and Department of Physics}}
\centerline{{\em Boston University, Boston, MA 02215, USA}}
\centerline{3) {\em Dipartimento di Fisica, Universit\'a di 
Roma ``La Sapienza''}}
\centerline{{\em Piazzale Aldo Moro 2, 00185 Roma, Italy}}
\date{~}
\medskip
\begin{abstract}
We present a detailed discussion of  a novel  
dynamical renormalization group scheme: the Dynamically Driven Renormalization 
Group (DDRG). This is a general renormalization method developed for 
dynamical systems with non-equilibrium critical steady-state.
The method is based on a real space renormalization scheme driven by 
a dynamical steady-state condition which acts as a 
feedback on the transformation equations. 
This approach has been applied
to open non-linear systems such as self-organized 
critical phenomena, and it  allows the analytical  
evaluation of scaling dimensions and critical exponents.  
Equilibrium models at the critical point can also be 
considered.  The explicit application to some models and the corresponding 
results are discussed.

\end{abstract}
\vspace{7 mm}

\newpage
\baselineskip 18pt
\section{Introduction}
The study of second order phase transitions introduced in the field of 
statistical physics the concept of criticality \cite{hua,par,yeo,domb}. 
A critical  
system is characterized by the absence of a characteristic lengthscale: 
the system 
fluctuates strongly and  correlation functions show 
non-analytic behavior. In the critical region therefore the usual tools of 
a physicist, principally perturbation theory, fail completely \cite{domb}. 
The renormalization group (RG) theory provides
a comprehensive understanding of second order 
phase transitions and critical phenomena \cite{am,ma,cres,bv}. This theory 
has been a major breakthrough in statistical physics and lead to the 
study of many others scale invariant and critical phenomena. Among these 
phase transitions associated with non-equilibrium states \cite{katz,zia}, 
fractal growth phenomena \cite{vic}, self-organized critical (SOC) systems 
\cite{soc,bak2} and a vast 
class of complex systems in which information spread over a wide range of 
length and time scales~\cite{bb}. 

Since both areas of critical and complex systems
deal with self-similar structures, it was natural for physicists who were 
familiar with RG techniques to consider these new problems as a possible 
playground for these methods. The situation turned out to be 
more complex, since the properties of non-equilibrium
critical phenomena are quite 
different from those of ordinary critical systems.
For instance, there is no ergodic principle  
and in general it is not possible to assign 
a Boltzmann weight to a configuration.
For the above reasons many authors concluded that these problems 
pose questions of new type for which it would be desirable to have a 
common theoretical framework \cite{kadt}. 
In the past decade an intense 
activity has been devoted to a better understanding of  these systems 
and theoretical methods are being developed~\cite{zia,fst}. 

Recently we introduced a renormalization 
scheme \cite{pvz} for sandpile models \cite{bak2}, 
that has later been  applied \cite{lpvz} to forest-fire 
models \cite{bak,ds}.
This approach deals with the critical properties of 
the system by introducing in the renormalization equations a dynamical 
steady state condition   
which provides the non-equilibrium stationary statistical weights to be 
used in the calculation.
In this way it is possible to characterize the fixed point dynamics and to 
compute analytically the critical exponents. 

Here we present the general formulation of this novel type of dynamical 
renormalization group: the Dynamically Driven Renormalization Group (DDRG),
a general theoretical method for dynamical non-equilibrium 
systems with critical stationary state. The essential idea is to combine a 
real space renormalization group (RSRG)
scheme with the dynamical steady-state condition which characterizes the 
stationary regime. The RG equations are driven by the steady 
state condition feedback from which we obtain the  
configurations approximate statistical weight to be used  in the 
dynamical renormalization of 
the master equation. 
While the approximate stationary distribution neglects correlations, these 
are considered in the dynamical renormalization. 
Finer scale correlations are then included in the approximate stationary 
distribution at the new scale which is calculated through the 
steady state condition  with renormalized 
dynamical parameters. 
This strategy gives an RG scheme which can be improved considering 
increasingly better approximations in both the dynamical renormalization 
and the stationary distribution. 

For the sake of clarity we discuss extensively the explicit application 
of the method to some driven dissipative systems referring to the 
present general framework. 
In particular we report in full details the results 
obtained for the critical height sandpile model and the forest fire model.
For this class of models the DDRG can be considered a general renormalization
scheme which provides a new class of analytical tools for the study 
of the stationary critical state.
The DDRG can also be  applied to ordinary dynamical critical phenomena 
for which 
the stationary state is characterized by Gibbs distributions. 
In this case the DDRG can be directly compared with other RG approaches.

The paper is organized as follows: sec.2 introduces the concept
of non-equilibrium steady-state and the approximate description in term 
of mean-field stationary conditions. Sec.3 presents in full detail 
the Dynamically Driven Renormalization Group. The basic recursion relations
are obtained and the conceptual scheme is discussed as well 
as the approximations
involved. Sec.4 shows the explicit application of the DDRG to sandpile 
and forest fire models. Sec.5 shows the application scheme 
to ordinary critical phenomena. 
Sec.6 is devoted to conclusions and perspectives.    
In the appendix A we discuss in detail
the coarse graining of time.

\section{Equilibrium and non-equilibrium  steady states}

The distribution over configuration space in equilibrium 
ensembles as well as many extended 
dissipative and nonlinear dynamical systems evolves in stationary 
states described by 
time independent probability distributions. 
The stationary state in some cases 
shows long range correlations and 
self-similar properties\footnote{In equilibrium ensembles 
this happens for special values
of the control parameter, i.e., at critical points}.
To describe these phenomena several models have been introduced:
here we will consider models 
defined on a discrete $d$-dimensional lattice. 
To each site of the lattice is associated a 
variable $\sigma_i$ that can assume 
$q$ different values ($\sigma_i=1, 2, 3,\cdots, q$).
The subscript $i$ labels the lattice site.  
A dynamics characterized by a set 
of parameters\footnote{ We should use a vector notation {
\boldmath $\mu$}$\equiv \{\mu\}$
to denote the set of all variables. 
For the sake of simplicity we use the simpler 
notation $\mu$.}
$\mu$ acts on these variables and 
defines the temporal evolution 
of the model.
The system can be described by $P(\sigma,t)$, 
the 
probability that at the instant $t$ the system is 
in the state $\sigma\equiv\{\sigma_i\}$. 
This is the usual way to characterize physical ensembles 
for which we want to know 
the statistical distribution in phase space. 

For stationary processes, the system 
is invariant under uniform time translations.
Thus, the variables $\sigma_i(t+\tau)$ are statistically 
indistinguishable from 
the untranslated variables $\sigma_i(t)$. This implies that all single time 
averages are constants and there is  a single time probability density 
$W(\sigma)$ independent of time. An ensemble of systems at thermodynamic 
equilibrium is, for example, stationary as well as  an ensemble in which the 
component systems are maintained in a non-equilibrium steady-state, at least 
after the ensemble has been sufficiently aged.
For equilibrium ensembles, the steady state statistical distribution is given 
by the Gibbs distribution.
In this case the functional dependence of the equilibrium distribution 
function 
$W(\sigma)$ on the parameters\footnote{For 
instance the temperature or the set of applied fields.} $\mu$
should be consistent with statistical mechanics and properly 
describe the static equilibrium properties of the system. 
In particular, the partition function and the 
equilibrium expectation values should show critical point singularities 
appropriate to the spatial dimensionality and the symmetry of the 
order parameter.
For non-equilibrium ensembles 
in principle one does not know how to assign a statistical 
weight to a given non-equilibrium configuration. An additional problem
is that, since the system is not in equilibrium, the distribution
changes in time.

One possibility is to study systems which have ``settled down'' into 
non-equilibrium steady states, so that the distributions, while non Gibbsian, 
have become stationary.
In this case we can describe,
at least approximately, 
the single-time statistical distribution \cite{kei},
since the densities $\rho_\kappa$ of sites in a particular 
state $\sigma_i=\kappa$ ($\kappa=1,\cdots,q)$  
do not change, on the average, as a function of time.
Associated to the set of time independent average densities
$\{\langle\rho_\kappa\rangle\}$ there is a unique 
stationary probability distribution, that characterizes 
single-time averages in the steady-state ensemble. 
We can therefore describe the average statistical 
state by means of stationarity
conditions for the system. These can be obtained from mean field 
equations of the form
\beq
\frac{\partial}{\partial t}\{\langle\rho_\kappa\rangle\}=
{\cal S}_\mu (\{\langle\rho_\kappa\rangle\})
\label{stat}
\eeq
where the operator ${\cal S}_\mu$  describes the evolution
of the system as a function of the dynamical parameters defined
above. 
For example, in the case of dilute gases the non-linear function that
gives the rate of change of the density becomes a non-linear functional, i.e.
the collision operator. In general, the operator 
${\cal S}_\mu$ represents
the sum of dissipative contributions, 
mechanical (non-dissipative) contribution, 
and the effect of other external fluxes. 
Note that the operator on the right-hand side of Eq.~(\ref{stat})   
is time independent, so that the differential equation is
first order in time. 

Time independent solutions of Eq.~(\ref{stat}) will be referred to as 
``steady-states'', although we should keep in mind that those are only the 
average states of the ensemble. 
Mathematically the steady-states $\langle\rho_\kappa\rangle$ are
determined by the equation
\beq
\frac{\partial}{\partial t}\{\langle\rho_\kappa\rangle\}={\cal S}_\mu 
(\{\langle\rho_\kappa\rangle\})=0
\label{stat2}
\eeq
The above equation can have more than one solution, even when  ${\cal S}_\mu$ 
is a linear function. In the following, however, we will consider only 
the presence of a single meaningful stationary state.
In ordinary statistical systems, the Eq.~\ref{stat2} represents the 
thermodynamic equilibrium condition. For driven dynamical systems, 
it describes the {\em driving} of
the system to the non-equilibrium steady state, 
by means of a balance condition.
We can express the stability or self-organization properties  
in mathematical form as
\beq
\lim_{t\to\infty}\langle\rho_\kappa (t)\rangle=\langle\rho_\kappa\rangle
\eeq
for any initial condition of the Eq.(\ref{stat}): the  
steady-state is an attractor for the dynamics.

It is possible to go further on these lines to find a more accurate  
description of the steady-state. 
For example, one could write down mean-field 
equations as Eq.(\ref{stat2}) for 
the average density correlations. We are, however, interested in systems which 
show {\em critical properties} in the stationary state. 
This implies non-analyticity and 
long-range correlations, 
so that it is impossible to go much further along with 
a mean-field description. 
To describe critical systems and scaling behavior, we have to turn 
our attention to renormalization group methods.

\section{The Dynamically Driven Renormalization Group}

In this section we present the general formalism of a new type of 
dynamical renormalization group especially suited for  
systems with non-equilibrium critical steady-state. 
In fact, by using the dynamical 
steady-state condition we are able to develop a renormalization strategy which 
allows us to compute critical exponents in a wide range of 
non-equilibrium systems. 
The method can also be applied to equilibrium dynamical critical 
phenomena for which the steady-state distribution is given by the 
equilibrium Gibbs distribution. 

In the following we will refer to RSRG schemes. 
The real-space formulation of the renormalization group, by virtue 
of its simplicity 
and versatility is a vital tool in the theoretical 
understanding of critical phenomena.
In contrast to momentum-space renormalization group, 
the RSRG, in general does not have any systematic  
way to treat the approximations: there is no small parameter 
that controls an expansion.
On the other hand, to improve the accuracy of the method 
one can use ``higher order''
techniques like the introduction of proliferation 
(additional couplings), larger cells, toroidal
or rectangular transformation or other extrapolation scheme 
\cite{cres,bv,rect}. In
this way, the critical exponents for several models have been obtained
with good precision. 

\subsection{Coarse graining and renormalization}

The essential ideas of the dynamic real space renormalization group approach
derive from Kadanoff's block analysis \cite{kad} and from 
coarse graining of time proposed 
by Suzuki \cite{suz}. We begin summarizing the main derivation of the method. 
We start with the following general Master Equation (ME):
\beq
\frac{\partial}{\partial t} P(\sigma , t)= {\cal L}(\mu) P(\sigma, t)
\label {lang}
\eeq 
where $P(\sigma,t)$ denotes the probability distribution function for the 
configurations of the system at time $t$, and ${\cal L}(\mu)$ 
is the temporal evolution operator, 
characterized by a set of dynamical parameters $\mu$. 
With the operator ${\cal R}$ we indicate the coarse graining operator, that 
eliminates degrees of freedom inside a cell and rescales time and space.
The application of ${\cal R}$ yields
\beq 
{\cal R} P(\sigma,t)= P^{\prime}(S,t')
\label{lang-rg}
\eeq
where $P^{\prime}(S,t')$ denotes the probability distribution 
for the coarse grained system. More explicitly, we can write:
\beq
{\cal R} P(\sigma,t)={\cal R}(e^{t{\cal L}} P(\sigma,0))= e^{t'{\cal L}'}
P^{\prime}(S,0)
\eeq
The scale transformation $t{\cal L}\to t'{\cal L}'$ yields the dynamical RG 
approach \cite{suz1}, while the scale transformation of $P(\sigma,0)$ 
corresponds to
the usual static RG approach \cite{ma,cres}. 
Denoting with the vector $\mu$ the 
parameters of the system, the RG yields 
recursion relations:
\beq 
\mu' = f(\mu) ~~~\mbox{ and }~~~ 
t'=g(t,\mu)\simeq t~g(\mu)
\label{recrel}
\eeq
from which it is possible to calculate the fixed points and the critical 
exponents of the model. These equations are obtained from the renormalization 
procedure which impose that $P^{\prime}(S,t')$ has 
the same functional form as 
$P(\sigma,t)$.

The most delicate problem in RSRG approaches is to take a 
partial elimination of degrees of freedom in 
Eq.s~(\ref{lang}) and (\ref{lang-rg}) \cite{bv}. 
To deal with this problem many approximate methods have been 
proposed \cite{suz1,nv}
and have been mostly applied to the kinetic Ising model \cite{glauber}.

Achaim and Kosterlitz \cite{A&K} perform a Migdal-Kadanoff \cite{mk} 
transformation on the
probability distribution function, assuming a functional form of the type:
\beq
P(\sigma,t)=e^{-H(\sigma)+h(t)}
\eeq
and treating the time dependent part as a perturbation. In this way they study
the relaxation of the probability density towards equilibrium. The dynamical
critical exponent is extracted from the scaling of the slower relaxation mode.
Suzuki et al.\cite{suz2} obtain the coarse graining of 
time studying the dynamical
equation for the first moment of the probability 
distribution (i.e. the magnetization).
The equation is decimated in the Migdal-Kadanoff approximation and the
equation of motion for the decimated modes are solved perturbatively.
The time scaling is chosen so that it keeps the equations in the same form.
A different formalism, suitable to study the properties
of the model even far from the critical region, was developed by 
Mazenko et al.\cite{mazinga} 
The coarse graining operator and the time rescaling are chosen
self-consistently in order to insure the Markoffian behavior of
the renormalized spin flip operator.
One then writes recursion relations for the two-point correlation
functions from which the critical properties of the model are extracted.

To develop a RSRG method suitable for irreversible 
non-equilibrium systems we will consider a more explicit treatment 
of the master equation. 
In particular we focus our attention on the 
dynamics of discrete models on a lattice characterized by a set of lattice 
variables $\sigma\equiv\{\sigma_i\}$, each of them being a $q$-state variable 
(see previous section), and by the 
dynamical parameters $\mu$.
In this case we can rewrite eq.(\ref{lang}) as 
\beq
P(\sigma,t)=\sum_{\{\sigma^0\}} \langle\sigma\mid 
T(\mu) \mid \sigma^0\rangle
P(\sigma^0,0)
\eeq
where $\langle\sigma\mid T(\mu) \mid \sigma^0\rangle$ 
is the transition probability 
from the configuration $\sigma^0\equiv\{\sigma_i^0\}$ to the configuration 
$\sigma\equiv\{\sigma_i\}$
in a unit time step $t$.
The symbol $\sum_{\{\sigma\}}$ will always mean a summation over all 
the configurations.
The operator $T$ is the discrete counterpart 
of the operator ${\cal L}$. 

We then coarse grain the 
system by rescaling lengths and time according to  the transformation 
$x\to bx$ and $t\to b^z t$. The renormalization transformation can be 
constructed through a renormalization operator  
${\cal R} (S,\sigma)$ that 
introduces the coarse grained variables set $S\equiv\{S_i\}$ 
and rescales the lengths 
of the system \cite{nv}. 
This operator contains all the information connecting the 
coarse-grained state with the original one. Not every choice 
of the operator $\cal R$ will lead to a meaningful transformation, 
and in constructing it one should be guided by physical insight. Moreover, 
the transformation should satisfy some general properties in order to respect 
the dimension and the symmetry of the internal 
space of the dynamical variables,
i.e. the renormalized variables should be of the same kind as the original. 
In general, $\cal R$ is a projection operator with the properties:
\beq
{\cal R}(S,\sigma)\geq 0 ~~~\mbox{ for any}~~~ 
S\equiv\{S_i\},\sigma\equiv\{\sigma_i\}
\eeq
and 
\beq
\sum_{\{S\}} {\cal R}(S,\sigma)=1 .
\eeq
These properties preserve the normalization condition
of the renormalized distribution. The explicit form of the operator $\cal R$
will be defined case by case in the various application of the method. 
Usually, it corresponds to a block transformation in which lattice 
sites are grouped together in a super-site 
that  defines the renormalized 
variables $S_i$ by means of a majority or spanning rule.

We subdivide the time step in intervals of the unitary time scale and
we apply repeatedly the dynamical operator $T$, obtaining
\beq
P(\sigma,N)=\sum_{\{\sigma^0\}}\langle\sigma\mid T^{N}(\mu) 
\mid \sigma^0\rangle P(\sigma^0,0)
\eeq
where $T^{N}$ denotes the application of the $T$ operator
$N$ times. We can therefore write the  eq(\ref{lang-rg}) for the coarse
graining of the system as follows:
\beq
P^{\prime}(S,t')=\sum_{\{\sigma\}}{\cal R}(S,\sigma)
\sum_{\{\sigma^0\}}\langle\sigma\mid T^{b^z}(\mu) \mid \sigma^0\rangle 
P(\sigma^0,0)
\label{start}
\eeq
where we have included the application of the operator 
$\cal R$ and $t'=b^z t$. The meaning
of $\langle\sigma\mid T^{b^z}(\mu) \mid \sigma^0\rangle$
has to be defined explicitly.
In the simplest case $b^z=N$ where  $N$ is an integer number.
Whenever is possible to define a continuous time evolution 
operator, $T^{b^z}$ can be found as an integration over time.
In general, since we are often dealing with discrete time 
evolution, we have to consider the $T^{b^z}$ as an effective 
dynamical operator. We will specify a projection operator ${\cal D}$ for 
the dynamics, that samples only the paths which lead to an
appropriate definition of the dynamical process at the coarse 
grained scale. Also in this case, as for the operator ${\cal R}$, spanning 
conditions or majority rules are obtained from physical considerations.
The projection operator is chosen 
in such a way to preserve the form of the operator 
$T$ at every scale. In this way it is possible 
to define recursion relations for 
the dynamical parameters. 
In appendix A, we present in detail the definition of the dynamical projector
operator and the explicit form of the effective operator $T^{b^z}$. 

In order to define the RG transformation more clearly, 
eq.~(\ref{start}) can be written as:
\beq
P^{\prime}(S,t')=\sum_{\{\sigma^0\}}\sum_{\{S^0\}}
{\cal R}(S^0,\sigma^0)
\sum_{\{\sigma\}}{\cal R}(S,\sigma)\langle
\sigma\mid T^{b^z}(\mu) \mid \sigma^0\rangle 
P(\sigma^0,0)
\eeq
where we used the properties $\sum_{\{S^0\}}{\cal R}(S^0,\sigma^0)=1$. 
By multiplying and dividing each term
by $P^{\prime}(S^0,0)=\sum_{\{\sigma^0\}}{\cal R}(S^0,\sigma^0)
P(\sigma^0,0)$,  and changing the order of summations we have 
\bea
\lefteqn{P^{\prime}(S,t')=}\nonumber \\
&\sum_{\{S^0\}}(\frac
{\sum_{\{\sigma^0\}}\sum_{\{\sigma\}}
{\cal R}(S^0,\sigma^0)
{\cal R}(S,\sigma)\langle\sigma\mid T^{b^z}(\mu) \mid \sigma^0\rangle 
P(\sigma^0,0)}
{\sum_{\{\sigma^0\}}{\cal R}(S^0,\sigma^0)
P(\sigma^0,0)})P^{\prime}(S^0,0)
\label{mefine}
\eea
which finally identifies  the renormalized dynamical operator $T'$ and 
yields the coarse grained master equation in the form 
\beq
P^{\prime}(S,t')=\sum_{\{S^0\}}
\langle S\mid T^{\prime}(\mu) \mid S^0\rangle P^{\prime}(S^0,0)
\label{meren}
\eeq
In the following, we apply this scheme to systems with a stationary 
distribution $P(\sigma,t\to\infty)=W(\sigma)$
given either by an equilibrium or a non-equilibrium 
steady-state. We can therefore study the renormalization of the dynamics 
for small deviations from the stationary state
\beq
P(\sigma,t)=W(\sigma)+\varepsilon h(\sigma,t)
\eeq
where $W(\sigma)$ is the steady-state probability distribution 
and $\varepsilon$ is an expansion parameter for the non-stationary part of
the distribution. 
In the lowest order approximation we can consider only 
the stationary part of the distribution in the renormalization of 
the evolution operator $T$, which, by comparing 
Eq.(\ref{mefine}) and (\ref{meren}), is 
given by:  
\bea
\lefteqn{\langle S\mid T^{\prime}(\mu) \mid S^0\rangle=}\nonumber \\
& &\frac
{\sum_{\{\sigma^0\}}\sum_{\{\sigma\}}
{\cal R}(S^0,\sigma^0)
{\cal R}(S,\sigma)\langle\sigma\mid T^{b^z}(\mu) \mid \sigma^0\rangle 
W(\sigma^0)}
{\sum_{\{\sigma^0\}}{\cal R}(S^0,\sigma^0)
W(\sigma^0)}.
\label{bas}
\eea
This is the basic renormalization equation that defines the 
dynamical evolution operator for the coarse grained system.
In principle one could also consider higher orders in $\varepsilon$ 
with an {\em Ansatz}\footnote{To first order in $\varepsilon$
one obtains:  
$$
h'(S,t')=
\sum_{\{S^0\}}(\langle S\mid T^{\prime}(\mu) \mid S^0\rangle_0 
h'(S^0,0)+\frac{\partial}{\partial\varepsilon}\langle S\mid T^{\prime}(\mu) 
\mid S^0\rangle_{\varepsilon=0}W'(S^0) 
$$
where the dependence of $T$ on $\varepsilon$ has been made explicit.
This equation describes the relaxation to the stationary state 
from a non stationary configuration.
Compare this  with \cite{A&K} where the relaxation to 
the equilibrium state after a perturbation (i.e. magnetic field)
was studied in a similar way.} for the form of $h$.
For the systems under considerations the zero order approximation will
give already non trivial results, since we are interested 
in the dynamics of the system once the steady state has been reached. 

To understand intuitively the above transformation (Eq.\ref{bas}) 
we can consider that
the operator ${\cal R}(S,\sigma)$ is a projection operator 
that selects only 
the configurations $\{\sigma_i\}$ which, at the coarse grained level, are 
mapped into the configuration $\{S_i\}$. The right-hand side term
of the above equation can be read as follows. The operator 
${\cal R}(S^0,\sigma^0)$
selects only the configurations which are renormalized in the starting 
configuration
$\{S^0_i\}$, each of them with a relative weight given by $W(\sigma^0)$ 
normalized with the factor $\sum {\cal R}(S^0,\sigma^0)W(\sigma^0)$.
For each of these configurations we compute the statistical weight of the 
evolution to the configuration $\{\sigma_i\}$,
through the paths selected by the dynamical operator $T^{b^z}$ 
(see the appendix).
Finally, we sum up only the contribution of the configurations
 $\{\sigma_i\}$ that renormalize
in the configuration $\{S_i\}$, selected from the operator 
${\cal R}(S,\sigma)$. In other words,
the new dynamical operator $T'$ is defined as the sum of the statistical
weight of all the paths that lead from any starting configuration 
that renormalizes 
in $\{S^0_i\}$, to any final configuration that renormalizes in  
$\{S_i\}$. Each path has 
then an additional weight given by the relative probability of 
the starting configuration 
$\{\sigma_i^0\}$. This last factor is the stationary statistical 
distribution, and preserve
the normalization of the transition probability matrix $T'$.
It is interesting to note that the stationary configurations 
distribution compares explicitly as a statistical weight in the 
renormalization equations.

The Eq.(\ref{bas}) is the basic renormalization equation from which 
the desired 
recursion relations are obtained. Imposing that the renormalized
operator $T'$ has the same form of the operator $T$, i.e. $T'(\mu)=T(\mu')$, 
we obtain the rescaled 
parameter set $\mu'=f(\mu)$. This implies that the renormalized  single time 
distribution $P^{\prime}(S,t')$ has the same functional form of the original 
distribution $P(\sigma,t)$. 
Since  we are dealing with discrete evolution operators $T$, we define 
the time scaling factor $b^z$ as the average number of steps we  
apply the operator $T$ in order to obtain that $T'(\mu)=T(\mu')$ for the 
coarse grained system. It therefore depend upon the 
parameters set $\mu$. This condition defines the time recursion 
relation $g(\mu)$, from which it is possible 
to calculate the dynamical critical exponent $z$ (see Eq.~\ref{recrel}). 

\subsection{Driving condition and recursion relations}

The scheme discussed so far is a general formulation valid 
for each system which 
exhibits a stationary state, and its application 
presupposes the knowledge of the 
explicit form of the steady-state distribution $W$. For example, 
in equilibrium phenomena $W$ is given by the Gibbs distribution.
In that case it is possible 
to apply several methods such as cumulant expansions and  exact or approximate 
decimation to obtain the form of the recursion relations.
For non-equilibrium dynamical systems in general we do not know the 
form of the steady-state distribution. 
We will therefore develop an approximate method to evaluate 
the stationary distribution to be used in the calculation of 
the renormalized master equation. 

The steady-state distribution can in general be split in two parts 
\beq
W(\sigma)=W^{\mbox{(i)}}(\sigma)+W^{\mbox{(c)}}(\sigma)
\label{decom}
\eeq
where $W^{\mbox{(i)}}(\sigma)$ and $W^{\mbox{(c)}}(\sigma)$ are, 
respectively, the incoherent and coherent part of the distribution. 
The incoherent part of the distribution has the property
\beq
\sum_{\{\sigma\}}\sigma_i \sigma_j W^{\mbox{(i)}}(\sigma)
=\bar{\sigma}^2.
\eeq
where $\bar{\sigma}$ is the single site average.
Hence, it does not include correlations among variables  
and expresses a mean field approximation for the system.
The coherent part $W^{\mbox{(c)}}(\sigma)$ can be  
subdivided in parts describing 
different kind of correlations: nearest-neighbors, 
next-nearest-neighbors
etc. The incoherent part is a factorized distribution,
that, for systems characterized by a q-state variables 
(see sect.2), has the form 
\beq
W^{\mbox{(i)}}(\sigma)= \prod_{i}\langle\rho_{\sigma_i}\rangle
\label{produ}
\eeq
where $\langle\rho_\kappa\rangle$ is the average density of sites in the 
$\kappa$-state. In this way, we have approximated the
probability of each configuration 
$\{\sigma_i\}$ as the product measure of the mean field  probability to 
have a state $\sigma_i$ in each corresponding site.
The incoherent part contribution to the renormalization equation is therefore 
particularly easy to obtain: for non-equilibrium 
steady-state we can use the stationarity condition, 
${\cal S}_\mu (\{\langle\rho_\kappa\rangle\})=0$, 
to evaluate the densities $\langle\rho_\kappa\rangle$. 
By solving the stationary condition equation, 
the average densities of the $\kappa$-states for the coarse grained system 
are obtained as a function of the dynamical 
parameters $\mu$ at the corresponding  iteration of the RG equations.
By inserting this approximate distribution 
in Eq.~(\ref{bas}), we get the following set of renormalization equations 
\bea
\lefteqn{\langle S\mid T(\mu') \mid S^0\rangle=}\nonumber \\
& &\frac
{\sum_{\{\sigma^0\}}\sum_{\{\sigma\}}
{\cal R}(S^0,\sigma^0)
{\cal R}(S,\sigma)\langle\sigma\mid T^{b^z}(\mu) \mid \sigma^0\rangle 
\prod_{i}\langle\rho_{\sigma_i^0}\rangle}
{\sum_{\{\sigma^0\}}{\cal R}(S^0,\sigma^0)
\prod_{i}\langle\rho_{\sigma_i^0}\rangle}
\label{ddrg1}
\eea
\beq
{\cal S}_{\mu}(\{\langle\rho_\kappa\rangle\})=0
\label{ddrg2}
\eeq
where the second equation denotes the dynamical steady state condition 
that allows the evaluation of 
the approximate stationary distribution 
at each coarse graining scale.
We call the Eq.~(\ref{ddrg2}) the {\em driving  condition}, 
since it drives 
the RG equations acting as a feedback on the scale transformation.

Rewriting these equations in the form of recursion relations and
adding the equation for the rescaling of time (see appendix A) we obtain
\footnote{Note that the $\mu$ recursion relation has to be read as 
a vectorial equation for the complete dynamical parameters set.}:
\beq
\mu'=f(\mu,\{\langle\rho_\kappa\rangle\})
\eeq
\beq
t'=t~g(\mu,\{\langle\rho_\kappa\rangle\})
\eeq
\beq
{\cal S}_{\mu}(\{\langle\rho_\kappa\rangle\})=0~~~~\Longrightarrow~~~
\langle\rho_\kappa\rangle=u_\kappa(\mu)
\eeq
in which  the driving condition appears explicitly.
The above set of equations (or, equivalently, the 
Eq.s~(\ref{ddrg1})-(\ref{ddrg2}) 
synthesizes the {\em Dynamically Driven Renormalization Group} at 
the lowest order. 
The fixed points $\mu^*=f(\mu^*)$  of the first equation govern
the critical behavior of the system.
The second equation gives the 
dynamical critical exponent $z$:
\beq
z=\frac{\log g(\mu^*)}{\log b}
\eeq
The third equation, the driving condition, can be seen as a 
feedback mechanism between the dynamical and stationary properties 
of the system. In addition, from the fixed point solutions
\beq
\langle\rho_\kappa\rangle^*=u_\kappa(\mu^*)
\eeq
the average stationary properties for the critical state are obtained.

In this form of the DDRG, we take into account only the uncorrelated part 
of the steady-state probability distribution. The results obtained 
are non trivial because correlations in the systems are considered in 
the dynamical renormalization of the operator $T$, that given a starting 
configuration traces all the possible paths leading to the renormalized 
final configuration. Moreover, geometrical correlations are 
treated by the operator $\cal R$ that maps the system by means of 
spanning conditions or majority rules. 
The renormalized uncorrelated part of the stationary distribution is  
evaluated from the stationary condition with renormalized parameters,
thus providing an effective treatment of correlations.

In principle we can also refine the method by including 
higher order contributions
to the unknown steady-state distribution. 
We have considered the simplest approximation for the driving condition,
that takes into account only the single point densities. It is possible
to consider also two points occupation probabilities, or even higher 
order clusters.
One would then write generalized mean field equations \cite{dick} for
the $n-$ point probability distributions and couple their solution to the
renormalization of the dynamical operator. In other words, at every iteration
of the scale transformation one should solve the generalized mean-field
equations with {\em renormalized} couplings $\mu$. This is in the spirit
of RSRG calculations where clusters of spins of higher order 
are introduced progressively in the calculations \cite{nv}. 

The DDRG scheme can be applied to both equilibrium and non-equilibrium 
critical systems, but it is particularly suitable to the latter case. 
In fact, the renormalization procedure does not act directly on the stationary
probability distribution, which is in general unknown
in non-equilibrium phenomena. 
The renormalization equations depend on
the stationary probability distribution only to weight local 
configurations, allowing the use of various 
approximations for its evaluation.
The relevant difference between our scheme and other dynamical 
RG methods is the fact that we obtain a set of equations which are 
independent on the specific form of the stationary probability distribution
of the system. 
This perspective is quite different 
from several previous real space 
dynamical RG approaches which were based on the 
explicit knowledge of the stationary distribution 
or the detailed balance hypothesis \cite{mazinga}.
The application range of these  methods was therefore restricted to 
to the relaxation dynamics of equilibrium systems.
 
Our goal is to describe far from equilibrium critical systems for 
which the Gibbsian equilibrium description is not valid. The method 
therefore finds potential application to the wide range of non-equilibrium
critical phenomena, such as driven-diffusive systems,
cellular automata and contact processes \cite{zia}. 
A real space renormalization treatment of these models
appears to be particularly suitable since they are usually
defined on a lattice with discrete time steps.
On the other hand, the method is limited by
the possibility to describe the system in terms of a reasonable 
mean field theory. In addition the present real space formulation 
is not easily implemented in systems that show non-local interactions.

In the next section we present the explicit implementation of the DDRG
to two specific cases. The purpose of the following section is 
to provide an example of how the method works in practice for non-equilibrium 
systems. 

\section{DDRG applied to sandpile and forest-fire models}

Many extended dissipative dynamical systems form structures 
with long range spatial and temporal correlations. 
The concept of self-organized criticality (SOC)
has been invoked by Bak, Tang and Wiesenfeld \cite{bak2} 
to describe the tendency of 
slowly driven systems to evolve spontaneously toward a critical 
stationary state with no characteristic time or length scale,
without the fine tuning of external parameters.
We are going to show that the slow driving condition is in fact
a fine tuning, which make the previous definition of SOC ambiguous.
A more appropriate definition of SOC, that takes into
account this observation, is given in \cite{grin}.

An example of SOC is provided by sandpile models: 
sand is added grain by grain on a
$d$-dimensional lattice until unstable sand (too large local slope of the 
pile) slides off. In this way the pile reaches a steady-state,
in which additional sand grains fall 
off the pile by avalanche events.
The steady-state is critical since avalanches of any size are observed.
This class of models can be used to describe a generic avalanche 
phenomenon, interpreting the sand as energy, mechanical stress or heat memory. 

The common characteristic of SOC systems is the presence of
a non-equilibrium critical steady-state, which
we can analyze using the DDRG formalism.
We studied the critical height sandpile automaton \cite{bak2} and 
the forest-fire model \cite{bak,ds}. 
These two models have been intensively studied numerically 
and can be considered as mile-stones in the field of SOC phenomena. 
In what follows we will show that the DDRG allows us to calculate 
analytically the critical 
exponents and to clarify the SOC nature of both models: we are able 
to study the fixed point and to identify the control and the order parameter
of the models. 

\subsection{Forest-fire model}

The Forest-Fire model (FFM)  has been 
introduced by Bak et al.\cite{bak} as a 
an example of SOC,
and has been then modified by 
Drossel and Schwabl \cite{ds}. 

\begin{figure}[htb]

\centerline{
        \epsfxsize=7.0cm
        \epsfbox{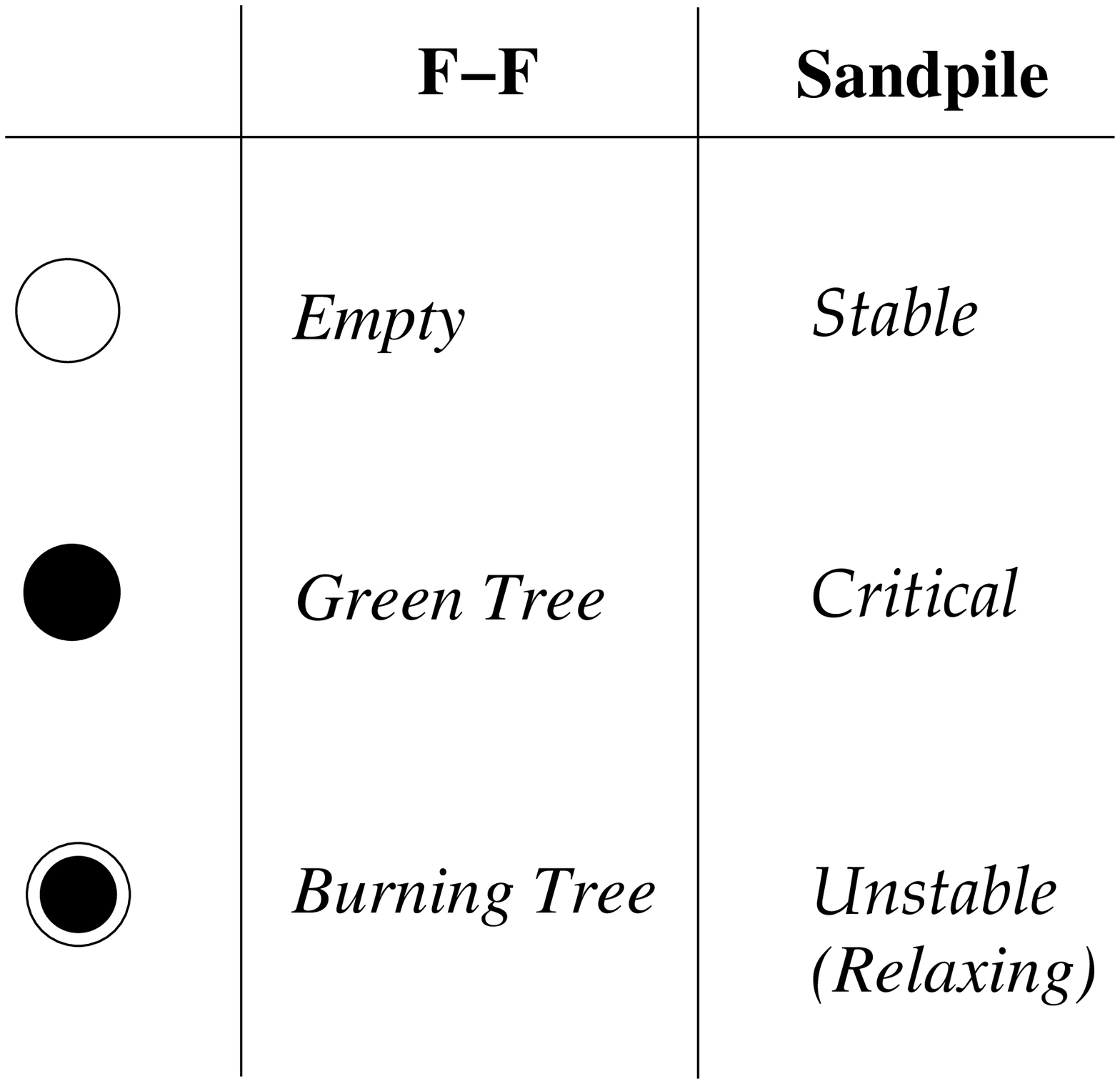}
        \vspace*{0.5cm}
}
\caption{~~}
\label{fig:1}
\end{figure}

The model is defined on a lattice in which
each site can be empty ($\sigma_i=0$), occupied by a green tree 
($\sigma_i=1$) 
or by a burning tree ($\sigma_i=2$) (see Fig~\ref{fig:1}). 
At each time step the lattice is updated as follows:

\begin{itemize}
\item[i)] a burning tree becomes an empty site;
\item [ii)]a green tree becomes a burning tree if at least one of its 
neighbors is burning;
\item [iii)] a tree can grow at an empty site with probability $p$;
\item [iv)]a tree without burning nearest neighbors becomes a burning 
tree with probability $f$.
\end{itemize}

The model was first studied in the case $f=0$ for the limit of very 
slow tree growth ($p\to 0$). In this limit the 
critical behavior is trivial: the model shows spiral-shaped fire fronts
separated by a diverging length $\xi\sim p^{-\nu_p}$, where
$\nu_p\simeq 1$ \cite{gras}.
In the case $f>0$,
the system was supposed to exhibit SOC
under the hypothesis of a double separation of time scales:
trees grow fast compared with the occurrence of lightnings  
and forest clusters burn down much faster than trees grow. 
This request is expressed by the double limit 
$\theta\equiv f/p \rightarrow 0$ and  $p \rightarrow 0$.
The critical state is characterized by a power law distribution 
$P(s)=s^{-\tau}$ of the forest clusters of $s$ sites 
(avalanches in the SOC terminology) and  the average  
cluster radius (the correlation length) scales as $R \sim \theta^{-\nu}$.  

In the past few years, a great amount of work has been done in order to 
describe the critical state of the Forest-Fire model and to calculate
the critical exponents. Numerical simulations \cite{gr2,cfo} 
show that in the time scale separation regime the model 
is close to a critical point with the avalanche critical
exponent $\tau$ given by $\tau \simeq 1$ in $d=1$ and $\tau \simeq
1.15$ in $d=2$. For the exponent $\nu$, describing the
divergence of the average cluster radius as $\theta \rightarrow 0$,
it has been found $\nu \simeq 1$ in $d=1$ and $\nu \simeq 0.58$
in $d=2$.
The one dimensional result has been recovered
exactly in \cite{ds2}. 
Simulations were performed also in higher dimensions \cite{cfo}. 
The critical dimension is believed to be $d=6$.

\begin{figure}[htb]

\centerline{
        \epsfxsize=7.0cm
        \epsfbox{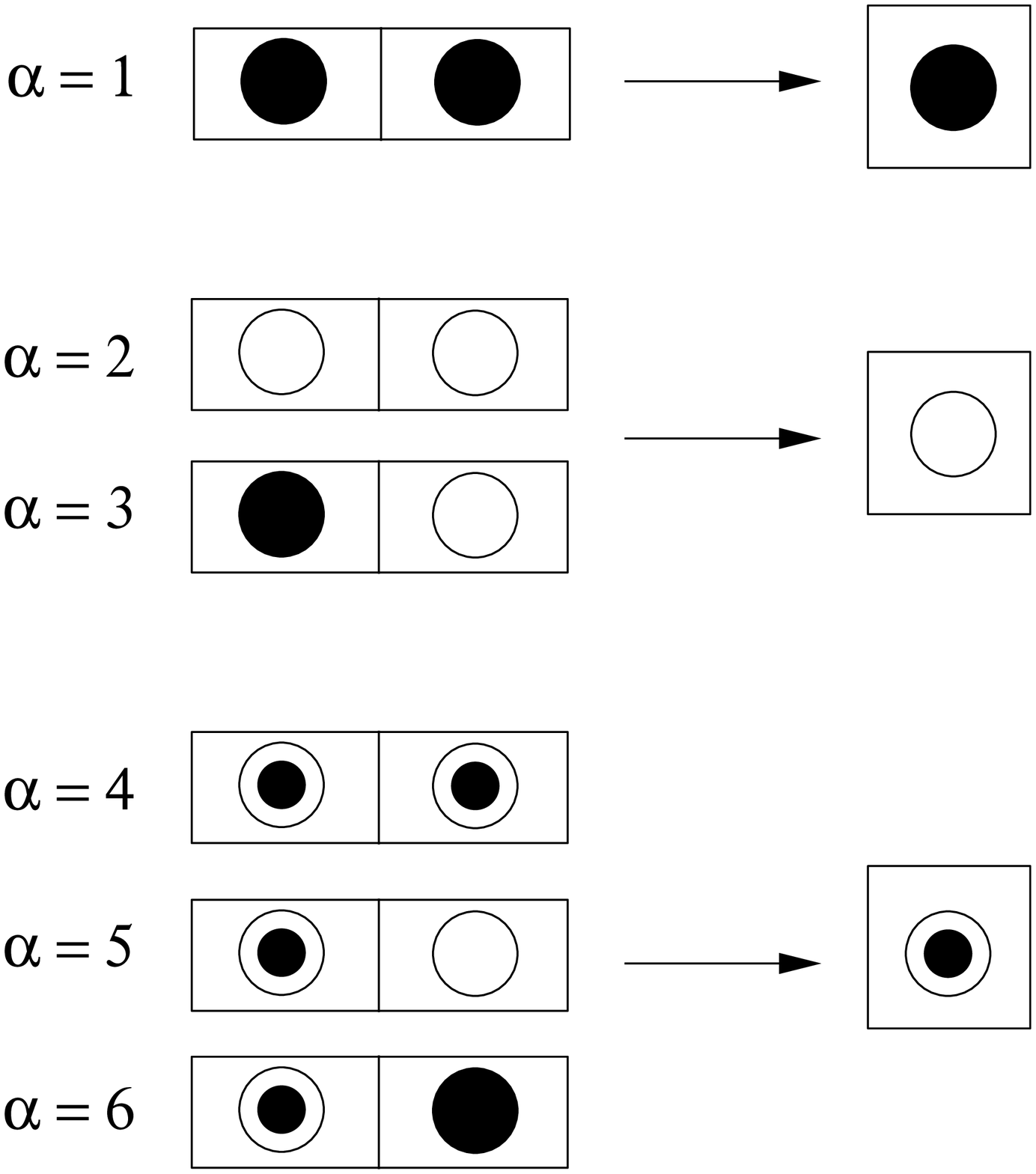}
        \vspace*{0.5cm}
}
\caption{~~}
\label{fig:2}
\end{figure}

To apply the DDRG to the FFM we follow step by step 
the strategy outlined in Sec.(3). 
For the sake of simplicity let us first
consider in full detail the one-dimensional
case. To define in practice the DDRG we first have to chose a form for 
the coarse graining operator. 
We use a cell-to-site transformation with scale factor $b=2$. 
In this case the operator ${\cal R}$ can be written in the following way:
\beq
{\cal R}(S,\sigma)=\prod_J{\cal R}(S_J, \{\sigma_i\}_J)
\eeq
where each term is acting on a specific cell $J$ and $\{\sigma_i\}_J$ 
denotes the configurations of spins belonging to that cell.  
Therefore, given a two sites cell, the operator ${\cal R}$ 
renormalizes it in a coarse grained site following only ``inside the cell''
rules. The rules defining ${\cal R}$ are as follows.
A two sites cell is renormalized as a tree site if it is spanned from left to 
right by a connected path of green sites. Accordingly, a cell is 
empty if it is not 
spanned by a connected path of green sites. Finally we consider a 
cell as burning if 
it contains at least one burning site. In this last case the spanning 
condition that ensures connectivity properties is not necessary 
because fire spreads 
automatically to nearest neighbor sites. The above renormalization 
prescription, 
i.e. the operator ${\cal R}$, is summarized in Fig.~\ref{fig:2}. 
We denote with an index $\alpha$ each two sites 
configuration $\{\sigma_i\}_J$ so that $\sum_{\{\sigma_i\}_J}\to 
\sum_{\alpha}$.
The dynamical rules of the FFM are local, therefore we can define  
matrix elements reduced to a single site\footnote{$\mid 0>$, 
$\mid 1>$ and $\mid 2>$ are  states in which the site $i$
is in the corresponding $\sigma_i$ state irrespective 
of the remaining of the system.}
for the dynamical operator $T$:
\beq 
\langle 0\mid T\mid 2\rangle = 1 
\eeq
\beq
\langle 0\mid T\mid 0\rangle = 1-p \mbox{\hspace{7mm}}; 
\mbox{\hspace{7mm}} \langle 1\mid T\mid 0\rangle = p
\eeq
\beq
\langle 1\mid T\mid 1\rangle = 1-f \mbox{\hspace{7mm}}; 
\mbox{\hspace{7mm}} \langle 2\mid T\mid 1\rangle = f
\eeq
In addition to the above rules,  
fire spreads between nearest neighbor sites.

The DDRG second step is the evaluation of the dynamical 
operator acting on the coarse grained variables via the 
the renormalization 
equations~(\ref{ddrg1},\ref{ddrg2}). We adopt a finite lattice 
truncation  on the two sites
cells subspace defined by the operator ${\cal R}$, and we obtain
the single site renormalized dynamical operator as 
\beq
\langle S_i\mid T'\mid S_i^0\rangle= \frac{\sum_{\alpha}
\sum_{\alpha'}
\langle\alpha'\mid T^{b^z}\mid \alpha\rangle W_{\alpha}}
{\sum_{\alpha} W_{\alpha}}
\eeq
where $\mid\alpha\rangle$ and $\mid\alpha'\rangle$ 
are the two sites cells states\footnote{For instance 
$\mid\alpha=1\rangle=\mid 1,1\rangle$ (see fig.2).}
which renormalize respectively 
in $\mid S_i^0\rangle$ and  $\mid S_i\rangle$. We keep the subscript $i$
for the latter states because they are referring to a single coarse
grained  site and not 
to a system's configuration. With $W_\alpha$ we denote the stationary
statistical weight of each $\alpha$ configuration. 

\begin{figure}[htb]

\centerline{
        \epsfxsize=7.0cm
        \epsfbox{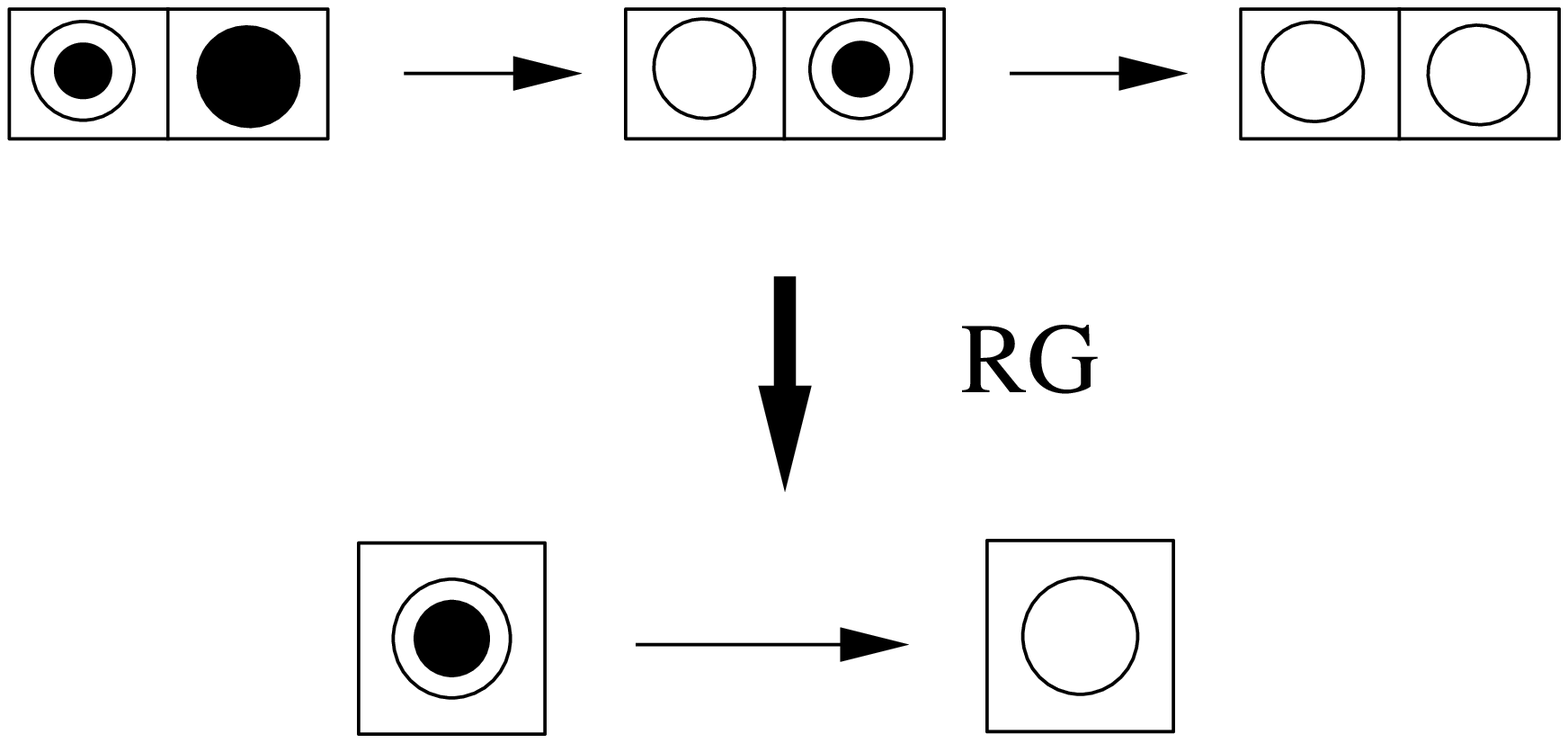}
        \vspace*{0.5cm}
        }
\vspace*{1cm}
\centerline{
\epsfxsize=7.0cm
        \epsfbox{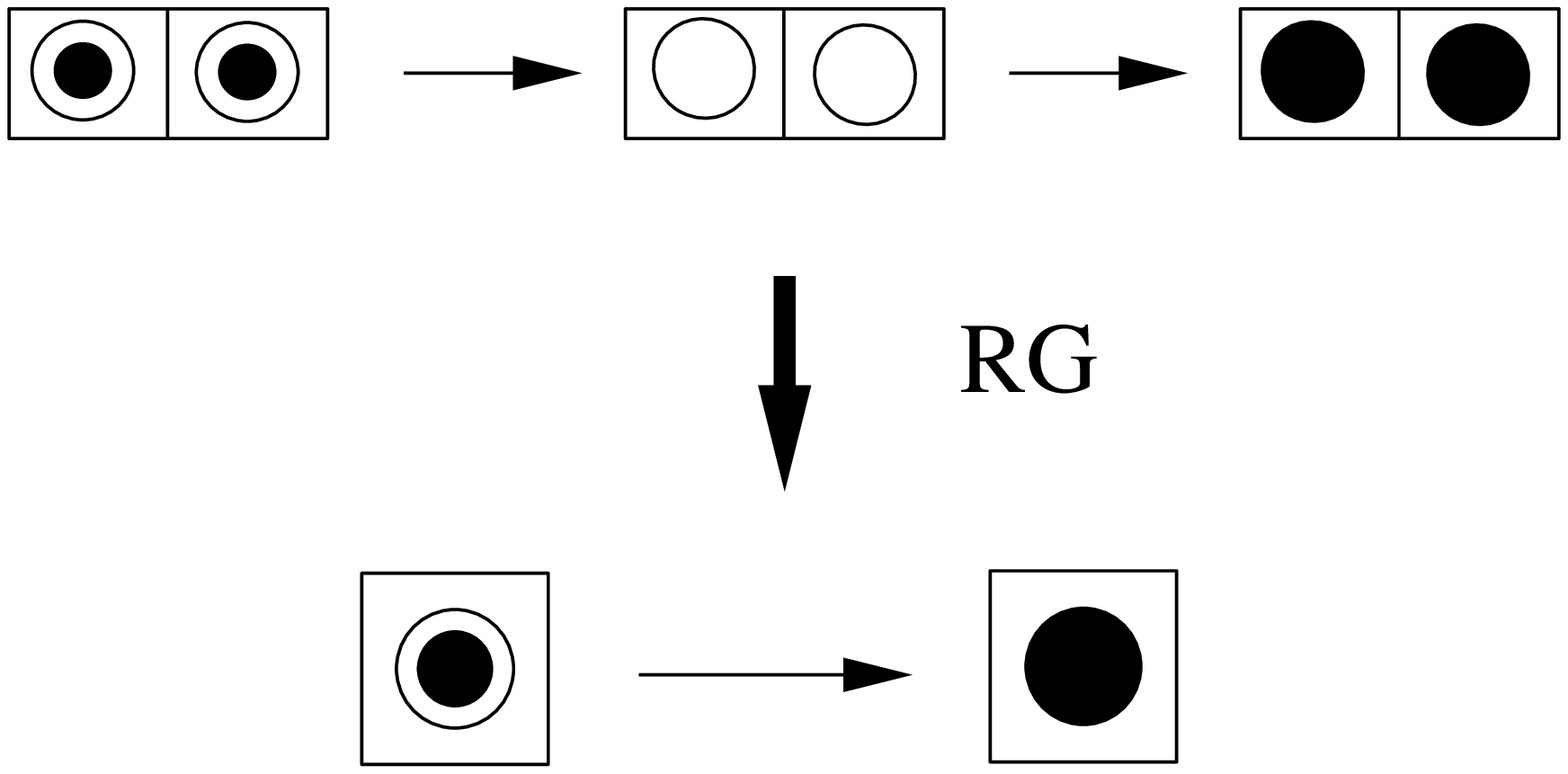}
        \vspace*{0.5cm}
}
\caption{~~}

\label{fig:3}
\end{figure}

Let us now evaluate explicitly the above equations.
Because we are interested in the critical behavior for $f\ll 1$ and $p\ll 1$ 
we can write non trivial RG equations keeping only terms up to the first 
order in $p$ and $f$.
In addition, in order to define consistently the recursion relations, 
the operator $T'$ 
must preserve its form at each scale; i.e. no proliferations are allowed. 
This implies that 
\beq
\langle S_i=0\mid T'\mid S_i^0=2\rangle= 1 + {\cal O}(p^2,f^2,pf)
\label{t02}
\eeq
where the higher order terms will be neglected in the recursion relations.
This implies that new parameters are not introduced in the description 
of the system. Thus,
to avoid proliferations in the burning event, we have to define a dynamical 
operator $T^{b^z}$ that leaves invariant Eq.~(\ref{t02}). 
It is easy to check that this is the case if $T^{b^z}=T^2$; i.e. $z=1$. 
In two time steps a burning cell evolves always in an empty 
one (Fig.~\ref{fig:3}a) if 
we neglect second order contributions in $p$ and $f$. 
In Fig.~\ref{fig:3}b we show 
a possible proliferation, which however has a weight $p^2$, and 
can therefore be neglected.
Avoiding proliferations in the burning event defines 
unambiguously the time scaling factor: the relevant time scale in the 
system is that of the burning process, as was already 
pointed out on the basis of 
numerical simulations \cite{ds}.

\begin{figure}[htb]

\centerline{
        \epsfxsize=7.0cm
        \epsfbox{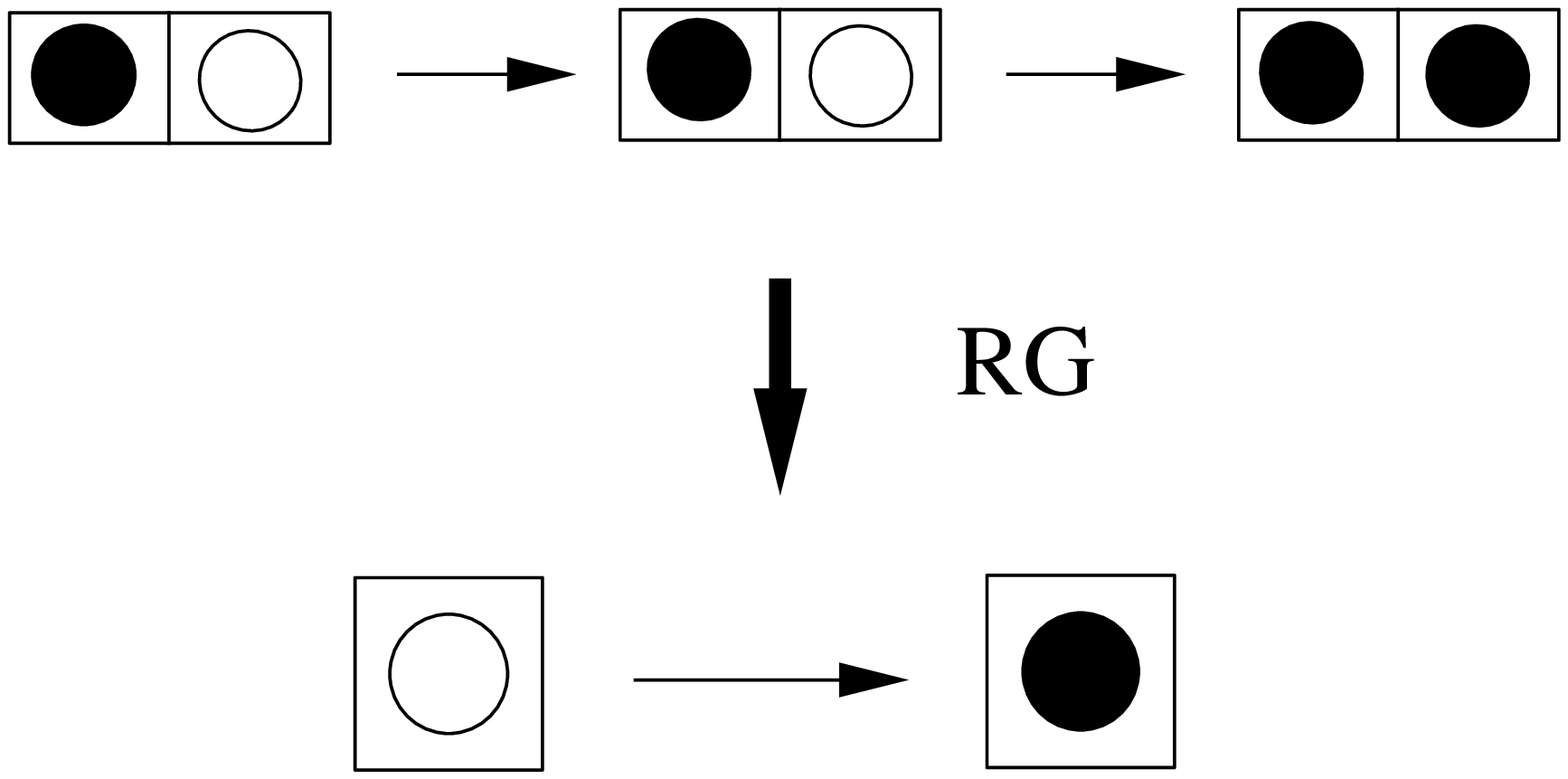}
        \vspace*{0.5cm}
}
\vspace*{1cm}
\centerline{
        \epsfxsize=7.0cm
        \epsfbox{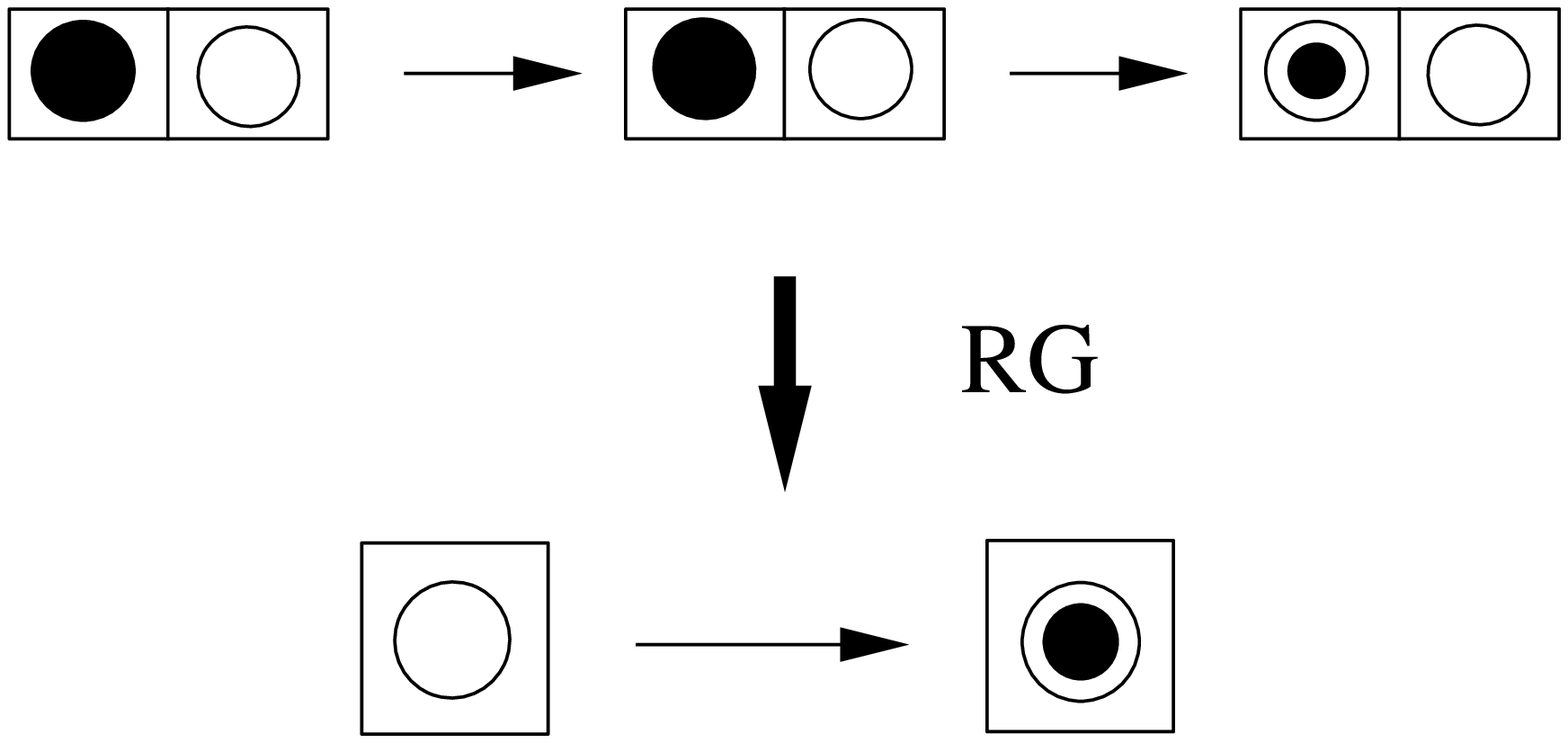}
        \vspace*{0.5cm}
}
\caption{~~}
\label{fig:4}
\end{figure}

The renormalization of the lightning probability in this 
framework is straightforward.
We have only one starting configuration, i.e. $\mid \alpha=1\rangle$ 
and the recursion relation is
\beq
f'=\langle S_i=2\mid T'\mid S_i^0=1\rangle=\sum_{\alpha'=4}^6 
\langle\alpha'\mid T^2\mid \alpha=1\rangle=
4f+ {\cal O}(p^2,f^2,pf)
\eeq
In the same way we obtain the expression for the 
renormalized growth probability $p$ as
\beq
\langle S_i=1\mid T'\mid S_i^0=0\rangle=
\frac{\sum_{\alpha=2}^3 \langle\alpha'=1\mid T^2\mid \alpha\rangle W_{\alpha}}
{\sum_{\alpha=2}^3 W_{\alpha}},
\eeq
from which follows
\beq
p'=2p\frac{W_3}{W_2+W_3} +{\cal O}(p^2,f^2,pf).
\eeq
A process that contributes to the above equations is 
shown in Fig.~\ref{fig:4}a. 
Finally, we have also to treat the normalization of the operator $T'$. 
In Fig.~\ref{fig:4}b we show a proliferation given by 
a process in which with probability $f$ an empty cell becomes a burning cell 
at the coarse grained level. In the previous equations, this last process 
is present only because of the normalization condition:
\beq
\langle S_i=1\mid T'\mid S_i^0=0\rangle=1 - 
2p\frac{W_3}{W_2+W_3} - f\frac{W_3}{W_2+W_3}
+{\cal O}(p^2,f^2,pf)
\label{prolf}
\eeq 
In order to avoid this proliferation we have  to impose a 
supplementary condition. If we restrict our analysis to the region
in which $f\ll p$, we can neglect terms linear in $f$ where terms linear in 
$p$ are present. This corresponds to truncate Eq.(\ref{prolf}) 
by keeping only the 
term linear in $p$, and thus eliminating the proliferation. 
The inclusion of
this proliferation would describe the behavior of the model in the limit
$f \simeq p$ which, from numerical simulations, 
is expected to be different. 
It is worth to remark that the above approximation 
corresponds to renormalize in a separate 
way the tree growth parameter $p$ and the lightning 
parameter $f$, assuming that they 
do not affect each other since they act on very different time scales. 

The steady state probability distribution $W_{\alpha}$ is approximate 
following the DDRG general scheme in the lowest order
\beq
W_{\alpha}=\prod_{i=1}^2\langle\rho_{\sigma_i}\rangle=n 
\langle\rho_{\sigma_1}\rangle\langle\rho_{\sigma_2}\rangle
\eeq
where $n$ takes into account the multiplicity due to 
symmetries of each configuration.

We can obtain the densities in the steady state 
from the following dynamical mean-field equations:
\begin{eqnarray}
\langle\rho_0(t+1)\rangle=(1-p)\langle\rho_0(t)
\rangle+\langle\rho_2(t)\rangle \\
\langle\rho_1(t+1)\rangle=\langle\rho_0(t)\rangle 
p+(1-f -(2d-1)\langle\rho_2(t)\rangle)\langle\rho_1(t)\rangle\\
\langle\rho_2(t+1)\rangle=\langle\rho_1(t)\rangle(f +(2d-1)
\langle\rho_2(t)\rangle)
\end{eqnarray}
where $d$ is the spatial dimension (see Ref.\cite{cfo} for their derivation). 
The long time limit ($t\to\infty$) solution of the 
above equations\footnote{It is worth to remark that also the
mean field equations are written for $f$ and $p$ close to zero.}  
provides the driving condition; i.e. the average steady state
densities.

Collecting all these equations we obtain the DDRG recursion
relations for the Forest Fire model, that in one dimension read as follows:
\beq
\left\{
\begin{array}{l}
p'= 2p\frac{\langle\rho_1\rangle}{\langle\rho_1\rangle+\frac{1}{2}\langle
\rho_0\rangle};\\
\theta'=2\theta\cdot 
\frac{\langle\rho_1\rangle+\frac{1}{2}\langle\rho_0\rangle}
{\langle\rho_1\rangle};\\
g(\theta,p,\rho)=2\\
\langle\rho_0\rangle-(1-\langle\rho_1\rangle)a/p=0;\\
\langle\rho_1\rangle-\frac{a}{\theta p+4\cdot a}-a\cdot 
\langle\rho_1\rangle =0 ;\\
\langle\rho_2\rangle-(1-\langle\rho_1\rangle)a=0.
\end{array}
\right.
\label{reneq}
\eeq
where we defined $a=p/(1+p)$. We
express the recursion relations in terms of the parameter $\theta$, in order
to compare with numerical simulation. 
It is important to emphasize again that the 
recursion equations (\ref{reneq}) are 
valid only in the double time scale separation $f\ll p\ll 1$
which defines the range of validity for our approximations. This limit
is the one for which the FFM shows non trivial critical behavior.

The flow diagram  
is stable with respect to different coarse graining rules,
and we find a repulsive fixed point in $\theta_c=0$ and $p_c=0$.
In order to discuss the critical behavior we have to 
linearize the recursion relations in the proximity of this fixed point and to 
find the relevant eigenvalues of the diagonal transformation:
\beq
\lambda_1=\left. \frac{\partial\theta'}{\partial\theta}\right|_{\theta_c,p_c}
\mbox{\hspace{5mm}}; \mbox{\hspace{5mm}}
\lambda_2=\left. \frac{\partial p'}{\partial p}\right|_{\theta_c,p_c}.
\eeq

The fixed point is repulsive, thus defining
the critical exponent $\nu$ in term of the largest eigenvalue $\lambda$ 
of the linearized renormalization equation 
$\nu={\log 2}/{\log(\lambda)}=1.0$.
>From simple scaling relations it is possible to obtain 
also the other critical exponents
which are summarized in table I. In this respect 
it is interesting to note that our method 
yields in the one dimensional case the exact results 
of the rigorous treatment of Ref.\cite{ds2}.
This is due to the relative simplicity of the one 
dimensional case, where the approximations 
involved - i.e. spanning conditions or proliferations - are irrelevant. 

In $d=2$ the calculation of the RG equations proceeds along 
the lines shown above \cite{lpvz}
but are algebraically more complex. In fact, one has to consider the
average over different paths, and new dynamical interactions are generated 
at each RG step. This is a signature that we need an approximation which
truncates the parameter space after each iteration so that it remains closed.
This is done by considering just the leading order in $f$ and $p$ in the 
renormalization equations, and ignoring any proliferations generated at each
group iteration.
With this scheme we obtain $z=1$, which is not an exact result also if in 
good agreement with numerical simulations ($z=1.04$ \cite{clar}). 
The fixed point 
In $d=2$ the fixed point remains $\theta_c=0$ and $p_c=0$, and
 the largest eigenvalue is given by 
$\lambda_1$, which determines the leading 
scaling exponent $\nu=\log b/\log \lambda_1=0.7$
(for $b=2$). The result is in good agreement
with numerical simulation ($\nu=0.6$ \cite{clar}). The
numerical value can be further improved 
by using larger cells \cite{lpvz}.

It is worth to remark that the DDRG allows to overcome the approximations 
present in the approach of Ref.\cite{lpvz}, where the time scaling was 
not properly considered because of the assumption of
an infinite time scale separation. In fact, 
in the limit $f=0$ the critical behavior is 
governed by the second eigenvalue $\lambda_2$. This eigenvalue 
and its relative exponent describes  the behavior of the correlation 
length in  the deterministic FFM. As opposed to $\lambda_1$,
the value of $\lambda_2$ depends on the absolute
value of the time scaling factor, and therefore could not
be obtained by  the scheme used in  Ref.\cite{lpvz}.
The numerical value we obtain in $d=1,2$ is 
$\nu_p=\log 2/\log \lambda_2=1.0$, which is in 
excellent agreement with 
the simulation results $\nu_p\simeq 1$ \cite{gras}.

In table I the results obtained in $d=2$
for $b=2$ are compared with numerical simulations. 
The existence of a  relevant 
scaling field and the general structure of the flow diagram is
stable with respect to different approximation schemes, and
more refined calculations lead systematically to an improvement 
in the numerical values of the results.
The FFM is critical along the line $\theta=0$ 
of the phase space, so that $\theta $ is equivalent to the reduced temperature 
in thermal phase transitions. In other words $\theta$ is the control parameter 
of the model, and the critical state is reached only by a fine tuning of 
$\theta$ to its critical value. 
The control parameter $\theta$ is the ratio between two very different
time scales, controlled by $f$ and $p$,  with a critical value fixed to zero.
In this situation, however, the
existence of a time scale separation makes the system very close to the
critical point without an apparent fine tuning of internal parameters.
Strictly speaking however, the system is
critical just in correspondence of the critical value of the control
parameter.

\subsection{Sandpile models}

Sandpile models are cellular automata \cite{bak2,zhang} 
defined in a $d-$ dimensional
lattice. A variable $E(i)$, that we denote by energy, is
associated with each lattice site $i$. At each time step
an input energy $\delta E$ is added to  a randomly chosen site.
When the energy on a site reaches a threshold value $E_c$ the
site relaxes transferring energy to the neighboring sites:
\beq
E(i)\to E(i)-\sum_{e}\Delta E(e) \nonumber
\eeq
\beq
E(i+e)\to E(i+e)+\Delta E(e)\label{din}
\eeq
where $e$ represent the unit vectors on the lattice. A typical choice
for the parameters is , for example, $E_c=4$ and $\Delta E(e)=\delta E=1$, 
but other possibilities have also been considered. 
The relaxation of the first site can induce a series of relaxations
generating an avalanche. Note that the energy is added to 
the system only when the configuration is stable (i.e. all the
sites are below the threshold).
The boundary conditions are usually chosen to be 
open so that energy can leave the system.
In these conditions the system organizes itself in to a stationary
state characterized by avalanches of all length scales. 
In particular the distribution for avalanches sizes $s$ decays
as a power law $P(s)\sim s^{-\tau}$, and the linear size of
the avalanche scales with time $r\sim t^z$.

This model has been extensively studied in the past by means of
numerical simulations \cite{manna,grasma,stella} and 
several exact results have been derived for  
Abelian sandpiles \cite{dhar}.

In the steady state each configuration of the system can be 
described by the energy probability distribution $W(\{E_i\})$. 
In the zero order approximation we consider the incoherent
part of the distribution:
\beq
W^{\mbox{(i)}}(\{E_i\})=\prod_i w(E_i)
\eeq
where $w(E)$ is the single site energy distribution.
The average  energy of a site evolves according to the following
equation, written in continuum notation:
\beq
\frac{dE(t)}{dt}=\delta E_{in}-\delta E_{out} 
\label{energy-drive}
\eeq
where $\delta E_{in}$ is the average energy entering into the site
either because of a relaxation in a neighboring site or
because of the external perturbations, and  $\delta E_{out}$
is the average energy dissipated by the site. 
At lower scale \newline $\delta E_{out}=
\langle\rho\rangle\sum_{e}\Delta E(e)$,
where we have defined 
\beq
\langle\rho\rangle=\int_{E_c-\delta E}^{E_c}w(E)dE
\eeq
as the probability that a site relaxes.

This equation suggest a simple way to describe the steady state
of the model. At any scale, we can divide the sites in {\em critical}
($\sigma_i=1$) and {\em stable} ($\sigma_i=0)$.
Stable sites do not relax
when energy is added to them. On the other hand critical sites 
relax when they receive an energy grain $\delta E_{in}$.  
In this formalism $\rho$ represent the density of critical sites.
For convenience we will also define unstable sites ($\sigma_i=2)$, 
as those that are relaxing, even though they are not present in the
static configurations of the system (see Fig.~\ref{fig:1}). 
These definitions can be
extended to a generic scale $b$. For instance, a cell at scale $b$ is
considered critical if the addition of energy $\delta E_{in}(b)$ induces a 
relaxation of the size of the cell (i.e. the avalanche
spans the cell).

The DDRG allows us to develop a general renormalization scheme 
for sandpiles which put in a broader and more systematic context 
the approach of Ref.\cite{pvz}.
To construct the DDRG recursion relations we have to describe the
relaxation at a coarse grained scale. We first note that 
the only non trivial matrix element of the time evolution operator is
the one describing the evolution of a cell from unstable  to stable.
This process occurs when a critical cell becomes unstable due to
the addition of energy. The inverse process, a stable cell becoming critical,
is simply due to the accumulation of energy and is not characterized
by critical exponents which, on the other hand, describe
the avalanche propagation. 

In a relaxation at the minimal scale energy is distributed equally in the
four directions. This is no longer the case at a coarse grained
level where different possibilities arise:   
the energy in principle can be distributed to one, two, three
or four neighbors. It is also worth to remark that in certain 
case unstable sites at the coarse grained scale  do not dissipate 
energy to nearest neighbors, representing just intra-site energy 
rearrangements. These processes defines the probability that 
relaxation events take place on the renormalized scale without 
energy transfer. All these events occur with probabilities
$$\vec{P}=(p_0,p_1,p_2,p_3,p_4)$$
In terms of the matrix element $\langle 0|T|2\rangle$ 
the vector $\vec{P}$ represents the probabilities
\beq
p_n=\langle 0|T|2\rangle_n
\eeq
where $\langle 0|T|2\rangle_n$ is the probability 
that a relaxing site  becomes stable and transfers
energy to $n$ neighbors.
In this way we have obtained the set of parameters  that
describes the dynamics. Of course the choice
of the parameters space is not uniquely determined, one encounters
proliferation problems typical of real space RG methods. 
For instance, higher orders proliferations are 
due to multiple relaxations
of the same site and sites becoming critical during the 
dynamical process (i.e.: $\langle 1|T|2\rangle$). In the following 
the practical implementation of the method considers just the 
minimal proliferation we reported above.

The renormalized matrix element
is then obtained by considering all the processes that span the cell
and transfer energy outside. This  rule implicitly
defines the effective dynamical operator $T^{b^z}$ (see App.A),
the renormalized parameters being:
\beq
p'_n =  \frac{\sum_{\{\sigma^0\}}\sum_{\{\sigma\}}
{\cal R}(S^0=2,\sigma^0)
{\cal R}(S=0,\sigma)\langle\sigma\mid T^{b^z} 
\mid \sigma^0\rangle_n 
W(\sigma^0)}{\sum_{\{\sigma^0\}}
{\cal R}(S^0=2,\sigma^0)
W(\sigma^0)}
\label{sandren}
\eeq
We proceed in defining explicitly a renormalization procedure for the  
dynamics by considering a finite truncation on four-sites cells. 
This corresponds to a cell-to-site transformation on a square lattice,
in which each cells at the coarser scale is formed by four 
sub-cells at the finer scale: the length scaling factor is $b=2$.
The operator ${\cal R}$ which define the coarse grained variables acts
on each specific cell through ``inside cell'' rules. A cell is renormalized 
as a relaxing one if it contains a relaxing sub-cell which transfers 
energy to a critical sub-cell. In this way we ensure that the occurring 
relaxation process is extending over the size of the renormalized length scale 
independently of the successive avalanche evolution. A critical cell is 
therefore defined by a cell which can be spanned by a path of relaxation 
events. The scheme  
considers only connected paths that span the cell 
from left to right or top to bottom. This spanning rule implies that only 
paths extending over the size of the resulting length scale contribute 
to the renormalized dynamics, and it ensures the connectivity 
properties of the avalanche in the renormalization procedure. 

Every cell at the coarser scale can be characterized by an index $\alpha$
that indicates the configuration  of  sub-cells, and we have that 
$\sum_{\{\sigma_i\}}\to\sum_{\alpha}$.  
The approximated stationary distribution (Eq.~(\ref{produ})) for 
each of these configurations is given by:
\beq
W_\alpha(\langle\rho\rangle)=
n_{\alpha}\prod_{i=1}^4\langle\rho_{\sigma_i}\rangle
\eeq
where $n_{\alpha}$ is a factor due to the multiplicity of each configuration.

By using this scheme and replacing  sums 
over configurations with sums over the index $\alpha$, the recursion 
relations can then be rewritten in the simpler 
form 
\beq
p'_n=\frac{1}{\cal N}\sum_{\alpha}
W_{\alpha}(\langle\rho\rangle)
\sum_{\alpha'}\langle\alpha'\mid T^{b^z}(p_{n'}) \mid \alpha\rangle _n
\label{recsand}
\eeq
where $\mid \alpha\rangle$, $\mid \alpha'\rangle$ denotes the 
four sites configurations which renormalize  
in $\mid S_i^0=2\rangle$ and  $\mid S_i=0\rangle$, respectively. 
In the above expression the denominator of eq.(\ref{sandren}) 
is adsorbed in the normalization factor ${\cal N}$. 

The time scaling factor and the explicit definition of the
effective dynamical operator $T^{b^z}$ can be found in the appendix A.
It is worth to remark that the present case is very different from the FFM
and the definition of the effective dynamical operator is non trivial.
The driving condition (eq.(\ref{ddrg2})) is obtained
from eq.(\ref{energy-drive})
by imposing stationarity. This implies that the stationary state is 
characterized by the balance between the energy that goes in and the 
energy that goes out of the system. We assume that energy is transferred
in ``quanta'' $\delta E = \delta E_{in}$ in each direction and we 
obtain on average 
\beq
\delta E =\langle\rho\rangle\delta E\sum_n np_n 
\eeq 
which implies
\beq
\langle\rho\rangle=\frac{1}{\sum_n np_n}
\label{balance}
\eeq
This relation gives the average density of critical 
sites in the steady-state,
allowing us to evaluate the approximate stationary 
distribution at each scale.
Therefore, eq.s(\ref{recsand}) and 
(\ref{balance}) are the complete 
set of the DDRG recursion relations.
The practical calculation of all the paths involved in the evaluation
of the above equation is very laborious and can be found 
elsewhere \cite{pvz,asa,iva}. moreover, we have shown a particular 
truncation scheme and more generally
the explicit evaluation of the recursion relations depends upon the 
chosen spanning condition and number of proliferations considered. 

In Ref.\cite{pvz}, it has been developed the simplest 
closed renormalization scheme which neglects in addition to the higher 
order proliferations mentioned above also 
the probability $p_0$. The flow diagram shows an attractive fixed point: 
the parameters evolves spontaneously towards their critical value. 
{}From the fixed point  the critical exponents can be computed. 
For a square cell of size $b=2$ the results obtained are  
$\tau=1.25$ and $z=1.17$,
in good agreement with computer simulations \cite{manna,grasma}.
The same method has also been applied to dissipative 
sandpile models \cite{pvz} and to directed sandpile models \cite{asa}.
The effect of the $p_0$ processes is being included and the results 
will be reported in a forthcoming paper. 
Recently, the expressions for the recursion relations
have been  linked to a branching mechanism that 
allows their calculation through a generating function. Using this method,
it is possible to include more proliferations to
the set of relaxation processes considered. The results obtained 
with this improved scheme allows an excellent qualitative 
and quantitative description of critical height sandpile model \cite{iva}.

It is worth to remark that in this scheme, as is usually done in computer
simulations, we are implicitly assuming a {\em slow driving} 
condition for the model. 
In fact, in the evaluation of the RG equations the external 
drive does not interfere with the relaxation process:  
dynamical processes during which sites become critical are not considered.
In order to overcome this approximation the energy flow on its turn should be 
renormalized and the addition of energy (possibility of new relaxation events)
during relaxation processes be allowed. Enlarging the phase space, 
a relevant parameter
appears, e.g. the driving rate. The driving rate is the 
incoming current of energy per unit time with respect to the total 
average energy in the system. Also in this case, as in the FFM, 
the system would be strictly critical 
just in correspondence of infinitesimal driving rate.
>From this point of view the only difference between Sandpile models
and Forest-Fire models is that the latter can not be studied in
a subspace with no relevant parameters without destroying the model itself.
The meaning of the self-organization is then related
to the widespread existence of systems with very different time scales and
not to the absence of relevant control parameters as often reported in
literature.

\section{DDRG and equilibrium critical phenomena}

The DDRG represents a general 
method to approach non-equilibrium critical systems with a stationary state 
and it allows also to study equilibrium models at the critical point.
In this last case the stationary state is characterized by 
probability densities written in terms of the Hamiltonian of the system
by means of the Gibbs distributions. This kind of systems have been 
extensively studied and well established theoretical tools are available to
approach them. The prototype of such systems is the Ising model.
On each site $i$ of some 
finite dimensional lattice we place a random variable $\sigma_i$ taking 
the values $\pm 1$. The Hamiltonian is
\begin{equation}
H( \sigma )= -J \sum_{\langle i,j\rangle}\sigma_i \sigma_j
\end{equation}
where the sum runs over all the nearest neighbors pairs. The 
correspondent Gibbs measure is
\begin{equation}
W(\sigma)= Z^{-1} \exp\left\{ - \beta 
H(\sigma)\right\}
\end{equation}
where $Z$ is a normalization factor (the partition function).

In order to apply the DDRG to equilibrium critical phenomena one has to  
treat these systems as dynamical systems with a well-defined stationary state.
Our discussion will refer to the Ising model with 
{\em heat bath} dynamics but the 
arguments can be generalized to different cases.
In a system described by a state $\left\{ \sigma_i \right\}$, 
the conditional probability to
flip a spin in a site $i$ is given by
\begin{equation}
\langle\sigma_i\vert T(\beta) \vert \sigma^0_i\rangle=
\frac{1}{1+e^{ \sigma_i \beta \left[J\sum_{nn}\sigma_j^0 \right]}} 
\label{heatbath}
\end{equation}
where $nn$ indicates the nearest-neighbors of the site $i$ 
and $\beta$ is the inverse of the temperature. 
The configuration $\left\{ \sigma_i \right\}$ is 
obtained from $\left\{ \sigma_i^0\right\}$ by flipping
the spin $i$. The heat-bath algorithm updates
$\sigma_i^0$ by choosing a new spin value $\sigma_{i}$, 
independently of
the old value of $\sigma_i^0$, 
from the conditional probability given by Eq.~(\ref{heatbath}).
All the others spins remain unchanged.
If $\left\{ \sigma_i \right\}$ and $\left\{ \sigma_i^0 \right\}$  are two 
arbitrary configurations of the system and 
$\langle\sigma\vert T(\beta) \vert \sigma^0\rangle$, 
$\langle\sigma^0  \vert T(\beta) \vert \sigma\rangle$ are the 
$1$-step transition probabilities between the two configuration,  
the following relation, known as detailed balance, is satisfied
by the heat-bath dynamics 
\begin{equation}
W(\sigma) 
\langle\sigma \vert T(\beta) \vert 
\sigma^0\rangle=
W(\sigma^0) 
\langle\sigma^0\vert T(\beta) \vert \sigma\rangle
\end{equation}
where $W(\sigma)$ and 
$W(\sigma^0)$ are the stationary (equilibrium) distributions
for  the states $\left\{ \sigma_i \right\}$ and 
$\left\{ \sigma_i^0 \right\}$.

>From the detailed balance we can deduce a stationarity 
condition for the dynamics which provides the steady-state 
distribution to be used in the DDRG.
For the actual implementation of the RG procedure one can use standard 
techniques for the definition of the operator $\cal R$, e.g. majority rules,
spanning conditions etc., and for the dynamical renormalization of the 
parameters $\mu$.
The work in this direction is still in progress and it will 
provide a test for the flexibility of our approach. 
In order to test the accuracy of the method we could compare 
the results for the probability densities obtained in our framework with 
those obtained renormalizing directly the Gibbs distributions with usual
RG schemes.

\section{Conclusions.}

In this paper we have presented a new renormalization scheme 
especially suited for 
systems with non-equilibrium critical steady-state. 
The essential idea of the method is the 
use of an approximate stationary probability distribution for the  
configurations of the system. This distribution is 
evaluated through a driving condition which identifies 
the single time averages in the 
steady-state. This approximate distribution is used as a weight in the  renormalization 
of the master equation which takes into account 
correlations due to the dynamical evolution. These 
correlation are thus considered in the 
approximate probability distribution of the renormalized system, 
which is calculated by the driving condition with 
renormalized dynamical parameters. 
The dynamical renormalization of the master equation is based on real space 
RG schemes which, in spite of their low systematicity, 
are very simple and intuitive. 
In addition, these schemes leave room for ``higher order'' 
techniques to improve the 
accuracy of the results.

The DDRG can be used to study systems with 
stationary critical state, but is 
particularly useful in non-equilibrium systems for which it is impossible 
to obtain the stationary probability distribution in the configuration space.
In fact, the DDRG appears to be extremely powerful in the study of SOC 
systems. These systems show a non-equilibrium steady-state very close 
to the critical point  for a wide range of internal parameters. 

The application to forest-fire and sandpile models shows
that the general results obtained are not affected by the approximations 
involved in our scheme, even though a more refined 
treatment of the calculation 
scheme, i.e. by introducing more proliferations, 
leads to an improvement of the values obtained for the critical exponents.
For SOC models the DDRG can be considered as a general theoretical 
framework that provides the analytical tools for a 
qualitative and quantitative 
study of the critical stationary state.

We considered as well the application to usual equilibrium phenomena, in  
the perspective of a better understanding of the method 
and its application to other systems. 
In particular the DDRG could
be very effective to study the critical properties of 
driven diffusive systems \cite{katz,zia}, which to our knowledge
have never been approached by real space RG methods. Work in
this direction is currently in progress.

\section*{Acknowledgments}

Part of the work described here was carried out in collaboration with 
L. Pietronero, and we are very grateful to him. A.V. has benefited
from stimulating discussions with and encouragements by J.M.J. van Leeuwen.
We are indebted with M. Vergassola for a technical remark.
We thank P. Bak, A. Ben Hur, R. Cafiero, 
 G. Caldarelli, P. Cizeau, R. Cuerno,
B. Drossel, S. Galluccio, P. Grassberger, E. V. Ivashkevich,
K. B. Lauritsen, M. Marsili, A.Stella 
and C. Tebaldi for useful discussions and comments.

\newpage

\appendix{\large \bf Appendix A}

\def\theequation{A. \arabic{equation}}
\setcounter{equation}{0}

The explicit evaluation of the time scaling factor 
and the corresponding recursion relation is a
complex task, because there is no standard
recipe for this in real space RG schemes.
Here we show a general strategy, 
which depends on the specific 
model for its actual implementation. 

The effective operator  $T^{b^z}$
contains all the dynamical processes that contribute 
to the definition of a meaningful renormalized
dynamics. We define the following transformation
\beq
\langle\sigma\mid T^{b^z}(\mu) \mid \sigma^0\rangle =
\sum_{N}{\cal D}_N\langle\sigma\mid T^{N}(\mu) \mid \sigma^0\rangle
\eeq
where ${\cal D}_N$ is the renormalization operator for the 
dynamical evolution of the system: it is a projection operator that  
samples only the paths of $N$ time steps which have
to be considered in the definition of the effective operator $T^{b^z}$.
To clarify  this point, let us consider for a moment
a spin flip dynamics in a Ising-like system. 
The matrix element $T(\mu)$ is non zero only for those configurations that 
are related by a single spin flip. In order to preserve the same
form for $T$ at a coarse grained scale, we have to impose that the
renormalized time evolution operator connects only configurations
that differ by a single coarse grained spin flip.
Flipping a coarse grained spin corresponds to the subsequent flipping
of different spins in the original system. The number of flipping necessary
to flip a macroscopic spin is not uniquely defined, but depend on the
configuration, both at the coarse grained and fine grained level,
and on the particular dynamical path chosen.
The effective operator $T^{b^z}$ is then a convolution 
of different $N$-steps operators.
 
The operator  ${\cal D}_N$ is chosen 
on the basis of physical considerations: spanning conditions etc. 
In addition, ${\cal D}_N$  should satisfy some general 
properties in order to preserve the symmetry
or the internal space of the dynamical variables. 
For instance, we have to ensure the normalization of the
effective dynamical operator by the property
\beq
\sum_{\{\sigma\}}\sum_{N}
{\cal D}_N\langle\sigma\mid T^{N}(\mu) \mid \sigma^0\rangle = 1
\eeq
Moreover, ${\cal D}_N$ 
must be consistent with the definition of the renormalization 
operator ${\cal R}$: it should describe
dynamical processes among renormalized 
variables of the same type of those 
given by the operator ${\cal R}$. Finally, 
${\cal D}_N$ has to preserve 
the form of the dynamical operator $T$ at each scale.
This condition imposes that the time scaling is
consistent with the length scaling used in ${\cal R}$. In this way it is 
possible to map the renormalized system in the old one with renormalized 
variables. 

As previously mentioned the operator ${\cal D}_N$ can assume a 
very simple form:
\beq 
{\cal D}_N= \delta_{N,N'}
\label{delta}
\eeq
where $N'= b^z$. In general, however, more complicated expressions are 
encountered (see Ref. \cite{pvz,lpvz}), since 
${\cal D}_N$ depends on the specific dynamics.
We have defined the effective evolution 
operator so that $T'(\mu)=T(\mu')$ for the renormalized 
system. On the other hand, the operator $T^{b^z}$ is 
in general the convolution of the discrete time step operator $T^N$  projected
by the renormalization operator ${\cal D}_N$. 
Thus, we can write it as a sum of terms that represent the statistical 
weight for the  evolution paths of $N$ time steps:
\beq
\langle\sigma\mid T^{b^z}(\mu) \mid \sigma^0\rangle =
\sum_{N} \gamma^N_{\sigma,\sigma^0}(\mu)
\label{gamma}
\eeq
These terms, which are obtained by the specific definition of the 
operator ${\cal D}_N$, can be used to calculate the time scaling 
factor $b^{z}=g(\mu)$ 
as an average over the renormalized dynamical processes. 
In fact, we can interpret the right sum in Eq.(\ref{gamma})
as an integral over dynamical evolution paths of different 
time duration; the time scaling factor being an average over these paths
whose statistical weights are the terms $\gamma^N_{\sigma,\sigma^0}(\mu)$.
For each case we have to find the process which defines the relevant 
time scale of the phenomenon. The time scaling factor will be 
the average over contributing 
paths  to the renormalization of this process.

To illustrate how the above procedure works in practice, we shall discuss 
explicitly the FFM and sandpile examples. 
For FFM we are in the simple case of Eq.(\ref{delta}). In fact the relevant 
process, i.e. the burning process, determines 
${\cal D}_N= \delta_{N,2}$. therefore only terms 
$\gamma^2_{\sigma,\sigma^0}(p,f)=\langle\sigma\mid T^{2}(p,f) 
\mid \sigma^0\rangle$ are allowed, so we simply have $b^{z}=g(p,f)=2$.
 
For the sandpile automata the calculation is rather laborious. In the slow 
driving regime, the only relevant time scaling length is given by the single 
relaxation event, for which we have at each scale 
\beq
\langle S_i=0\mid T^{\prime} \mid S_i^0=2\rangle = 1
\eeq
The operator ${\cal D}_N$ selects those spanning paths that lead to 
a relaxation
process at the coarser scale.
By using the cell-to site transformation defined in Sec.4.2,
it is easy to show 
\beq
\sum_{\{\sigma\}}
{\cal R}(S_i^0=2,\sigma^0)
{\cal R}(S_i=0,\sigma)\sum_N\gamma^N_{\sigma,\sigma^0}(p_{n'})=
\sum_{\alpha'}\sum_N{\cal D}_N\langle\alpha'\mid T^{N}(p_{n'}) 
\mid \alpha\rangle
\label{gamsand}
\eeq
where $\mid \alpha\rangle$, $\mid \alpha'\rangle$ denotes the 
four sites configurations which renormalize  
in $\mid S_i^0=2\rangle$ and  $\mid S_i=0\rangle$, respectively. 
The sum over $n$ denotes we are considering relaxation processes
without distinguishing the number of affected nearest neighbors.
The operator ${\cal D}_N$ is therefore defined explicitly as an operator 
acting on the paths internal to four sites cells. It selects for each $N$
just relaxation paths which consist of $N$ connected non-contemporary
relaxation events that leave the cell without unstable sites. In a 
mathematical forms it reads as 
\beq
{\cal D}_N=\prod_{i\in\{ \alpha'\}}(1-\delta_{2,\sigma_i})
 \prod_{J=0}^{N-1}\sum_{m=1}^{4}\delta(m-\sum_{i\in \{\alpha_J\}}
\delta_{2,\sigma_i})
\eeq
where $\alpha_J$'s are the intermediate cell configurations during 
the dynamical evolution and 
$\sum_{i\in \{\alpha_J\}}$ denotes the sum over all the sites in the 
cells. In the above expression, each delta function acts on a different
intermediate cell eliminating those paths which do not have activity 
at each dynamical step. Furthermore, the operator ensures that in the 
cell $\alpha'$ ($N$th step) no activity is present; i.e the process has 
stopped.
Finally we have to write the equation that gives the time scaling factor 
from the total average over contributing processes to the renormalized 
matrix element $\langle 0\mid T^{\prime} \mid 2\rangle$ :
\beq
g(\mu)=
\frac
{\sum_{\{\sigma^0\}}\sum_{\{\sigma\}}
{\cal R}(S^0_i=2,\sigma^0)
{\cal R}(S_i=0,\sigma)\sum_N N \gamma^N_{\sigma,\sigma^0}(p_{n})
W(\sigma^0)}
{\sum_{\{\sigma^0\}}{\cal R}(S^0_i=2,\sigma^0)
W(\sigma^0)},
\label{tscal}
\eeq
and by inserting  Eq.(\ref{gamsand}) in the above relation 
we finally obtain 
\beq
g(p_{n})=\frac{1}{\cal N}\sum_{\alpha} W_{\alpha}(\langle\rho\rangle)
\sum_{\alpha'}\sum_N N{\cal D}_N\langle\alpha'\mid T^{N}(p_{n'}) 
\mid \alpha\rangle
\eeq 
where we used the DDRG scheme to explicitly get the stationary weights
and $\cal N$ is an opportune normalization factor. 

The above relations will provide the consistent 
rescaling of time by imposing that $b^z=g(p_n^*)$ from which it is possible 
to calculate the dynamical critical exponent.
This also shows that in general the factor 
$g(\mu)$ is a function of the 
dynamical parameters.

\newpage

{\bf \large FIGURE CAPTIONS}

\begin{itemize}

\item{\bf Fig.1} The sites classification in the Forest Fire
and in the sandpile model.

\item{\bf Fig.2} The rules defining the renormalization operator $\cal{R}$
for the one dimensional Forest Fire model.

\item{\bf Fig.3} (a) Evolution of a burning cell into a
empty cell in two time steps. 
(b) A possible proliferation in which a burning
cell evolves in two time steps into a green cell. This process
is of order $p^2$ and can therefore be neglected. 

\item{\bf Fig.4} (a) A process contributing to renormalization of $p$:
an empty cell becomes green due to the growth of one site.
(b) A possible proliferation of the growth dynamics in which
an empty cell becomes burning. In the limit $f\ll p$ this
process can be neglected.

\end{itemize}

\newpage

{\bf \large TABLE CAPTIONS}

\begin{itemize}

\item{\bf Table I} In this table we summarize our results 
for the Forest Fire Model critical exponents obtained with different
approximation schemes. For comparison we report also the 
exact or numerical results.
$^*$: Exact results from \protect\cite{ds2}.
$^+$: Numerical results from \protect\cite{gr2,clar}.

\end{itemize}

\newpage

\centerline{TABLE I}

\begin{table} 
\centering 
\begin{tabular}{|c|c|c|c|}
\hline $d=1$ & $\nu$ & $z$ & $\tau$ \\\hline \hline 
RG & $1.0$ & $1.0$ & $1.0$ \\\hline
Exact results$^*$ & $1.0$ & $1.0$ & $1.0$\\\hline
\hline 
\hline $d=2$ & $\nu$ & $z$ & $\tau$ \\\hline \hline 
RG 2 x 2 & $0.73$ & $1.0$ & $1.19$ \\ \hline
Numerical results$^+$ & $0.58$ & $1.04$ & $1.15$ \\ 
\hline \hline
\end{tabular}

\end{table}
\newpage

\newpage

{\bf FIGURES}

\vspace{3cm}

\end{document}